\documentclass[tightenlines,aps,prl,twocolumn,showpacs]{revtex4}
\usepackage{mathrsfs, mathbbol}
\usepackage{graphicx}
\usepackage{eufrak}
 \usepackage{verbatim}
\newcommand{\ket}[1]{| #1 \: \rangle}
\newcommand{\bra}[1]{\langle \: #1 |}
\newcommand{\out}[2]{{\ket{#1}\bra{#2}}}

\newcommand{\der}[2]{\frac{{\rm d}  #1}{{\rm d}  #2}}
\newcommand{\bea}{\begin{eqnarray}}
\newcommand{\eea}{\end{eqnarray}}

\newcommand{\lr}[1]{ \left( #1 \right)}
\newcommand{\com}[2]{ \left[ #1,#2 \right]}
\newcommand{\acom}[2]{ \{ #1,#2 \}}

\def \r{\rho}
\def \k{\kappa}
\def \a{\alpha}

\def \s{\sigma}
\def \o{\omega}
\def \O{\Omega}
\def \ah{\hat{a}}
\def \ahd{\hat{a}^{\dagger}}

\begin{document}

\title{Periodic revival of entanglement of two strongly driven qubits in a dissipative cavity}

\author{Marcin Dukalski}
\author{Ya. M. Blanter}
\affiliation{Kavli Institute of Nanoscience, Delft University of Technology, Lorentzweg 1, 2628 CJ Delft, The Netherlands}
\begin{abstract}
We study the dynamics and decoherence of a system of two strongly driven qubits  in a dissipative cavity. The two qubits have no direct interaction and are individually off-resonantly coupled  to a single mode of quantized radiation. We derive analytical solutions to the Lindblad-type master equation and study the evolution of the  entanglement of this system. We show that with non-zero detuning between the quantum and classical fields,  the initial decay of the entanglement is followed by its revival periodic in time. We  show that different Bell states follow  evolutions with different rates.
\end{abstract}

\maketitle

Entanglement is a resource essential for successful implementation of quantum information processing \cite{Nielsen}, and engineering sustainable entanglement is a necessary requirement for any physical realization of a quantum computer. Since a number of solid-state qubits have been successfully realized and
 single qubit operations have been demonstrated \cite{NazarovBlanterbook}, the attention of experimentalists is now mainly focused on interaction between qubits.
It is therefore of vital importantce to study processes which can generate or alter entanglement through an interaction between two or more qubits.

In many realizations it is easier to couple solid-state qubits via an optical or a microwave cavity rather than directly. Entanglement between qubits coupled to optical cavities has been studied in a very broad context. It is known that dissipation, both in qubits and in the cavity, can very quickly lead to disentanglement of two qubits \cite{J.H.Eberly04272007, PhysRevB.66.193306, PhysRevLett.93.140404}. This process, known as the entanglement sudden death (ESD), was confirmed experimentally \cite{PhysRevLett.99.180504, M.P.Almeida04272007}. For carefully chosen initial conditions and system parameters, ESD can be followed by the entanglement revival -- entanglement sudden birth \cite{ESB1, ESB2,  ESB3}, which, however, was found to never reach the initial (before ESD) degree of entanglement. Later it was shown that two directly interacting qubits can become entangled via a spontaneous decay \cite{TanasFicek}. Numerical results supporting the claim that under spontaneous atomic decay some atom-field entanglement can be generated have been presented in Ref. \cite{Casa2008}.

In this Letter, we consider two qubits which are strongly coupled via a dissipative cavity, so that the direct interaction between the qubits is negligible. The qubits are driven by a classical ac off-resonant field. As a result of detuning between the classical pumping frequency and the quantum cavity eigenmode, we find a novel entanglement behaviour where two initially entangled qubits experience periodic entanglement drops and revivals. Their concurrence, despite its initial sudden decrease with time, asymptotically tends to its initial value. Moreover, we find that once the qubits are unevenly coupled to the cavity, the decoherence free subspaces previously discussed \cite{PhysRevA.77.033839} disappear and the previously unaffected states also decay however now at slower decay rates. Our results can be applied to superconducting qubits coupled to a microwave cavity or to an NV center in diamond strongly coupled to an optical cavity.

Below, we first present the derivation of the effective multi-qubit Hamiltonian in the interaction picture. Then we find the solutions to a single qubit interacting with a coherent mode of radiation in a dissipative cavity, which is followed by an extension of this problem to the two-qubit case. Finally we analyze the temporal dynamics of entanglement of this system.
\begin{figure}
	\centering
		\includegraphics[width=8cm]{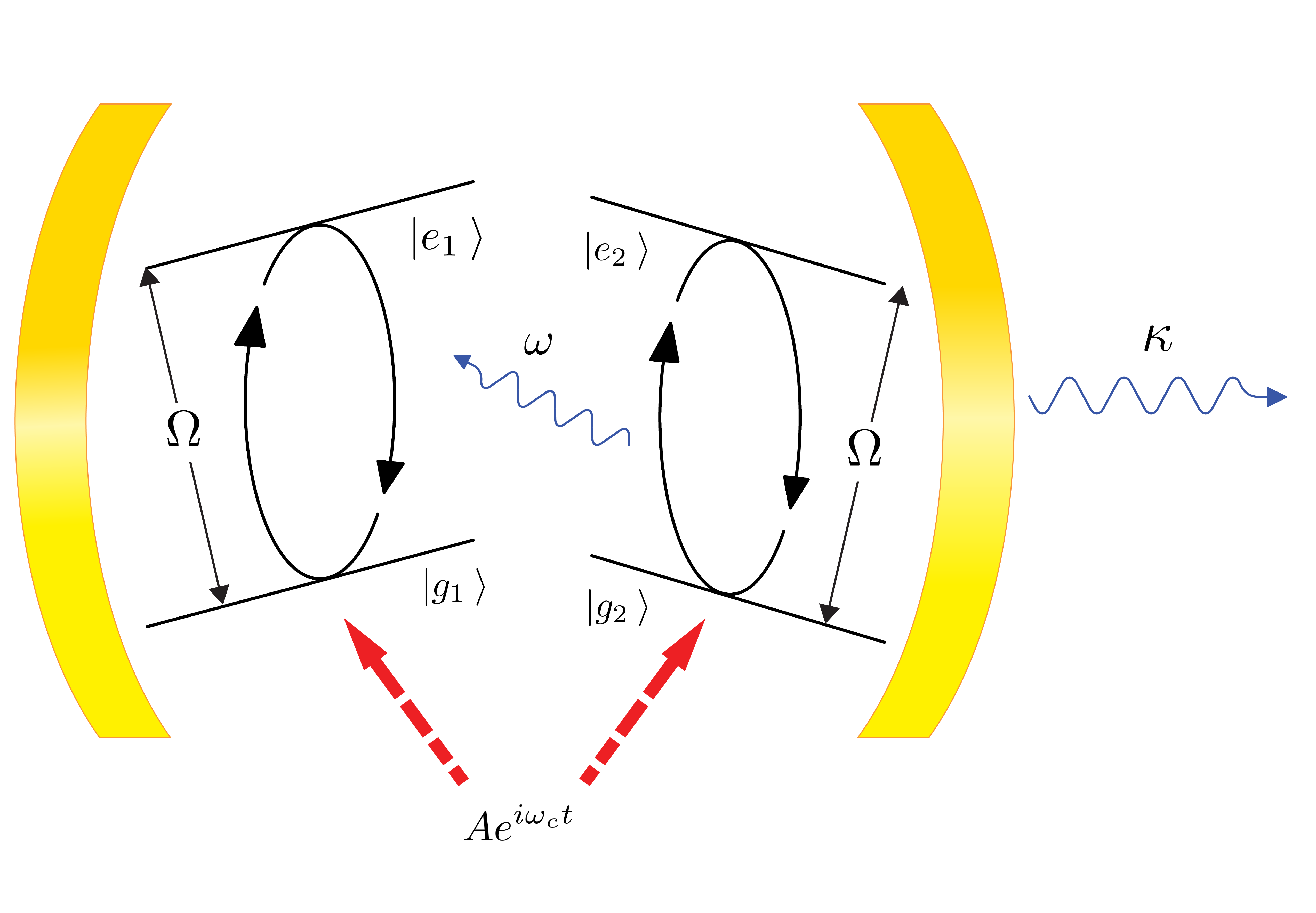}
	\caption{Schematic representation of the setup.}
	\label{fig:cavity}
\end{figure}

{\em The model}. The interaction of a qubit and a cavity is commonly described in terms of the Jaynes-Cummings Model (JCM) \cite{JC} which is one of the few interacting quantum systems admitting closed form solutions. JCM and its several variants have become a textbook tool to discuss coupled qubit and photon systems. Recently, it has been realised, that the qubit-field interaction with an additional strong driving also can be solved analytically \cite{1993OptCo..99..344J, PhysRevA.69.023812, 2003PhRvL..90b7903S} even if the cavity dissipation is also included in the system. Moreover, in Refs. \cite{Bina2008, Bina2009, Bina2010, ZhangXu} it has been proven that the solutions to the equations of motion for strongly driven qubits interacting through a cavity vacuum field can be extended to an unlimited number of qubits, which can not be achieved in the simple JCM Hamiltonian.

To keep the treatment general, we consider a Hamiltonian of a system of $N$ identical qubits coupled to a single-mode cavity and additionally driven by a classical electromagnetic field \cite{2003PhRvL..90b7903S},
\bea
\hat{H}&=&\frac{\O}{2} \sum\limits_{j=1}^N   \s^z_j +   \o \ahd\ah+A\sum \limits _{j=1}^N \lr{e^{-i \o_c t} \s^+_j+e^{i \o_c t} \s^-_j}\nonumber\\
&~&~~~~~~~~~+  \sum\limits_{j=i} g_j\lr{\s^+_j \ah+\s^-_j \ahd} \, ,
\eea
where $\O $ is the level spacing of the qubits, $\o$ is the frequency of the eigenmode of the cavity, $A$ and $\o_c$ are the amplitude and the frequency of the classical field, and $g_j$ is the coupling strength between the $j^{\rm th}$ qubit and the cavity mode. In addition to that $\s^z$ and $\s^{\pm}=\frac{1}{2}\lr{\s^x\pm i \s^y}$ are the (linear combinations of) Pauli matrices and $\ah$ $\lr{\ahd}$ is the annihilation (creation) operator of the quantum field mode. Throughout the Letter we set $\hbar=1$.

We assume that the qubits are driven strongly and that they are very stable and moderately coupled to the cavity mode, $A\gg \o,\delta \gg g \gg \gamma$, where $\gamma$ stands for the qubit decay rates. Therefore, we can ignore the qubit dephasing or decoherence rates as well as the energy violating ("counterrotating") Rabi Hamiltonian terms \cite{2003PhRvL..90b7903S}, $\s^+ \ahd$ and $\s^- \ah$. Additionally, we take the classical field to be sufficiently off-resonant, $\o_c\neq \o$, so that we can ignore  the classical field-cavity coupling. 

We begin by  applying an entanglement preserving  time-local unitary transformation
\bea
\hat{H}&\to& \hat{H'}=\hat{U}^{\dagger}\hat{H}\hat{U} -i\hat{U}^{\dagger}\partial_t \hat{U}  \, ,  \nonumber\\
\ket{\psi}&\to& \ket{\psi'}= \hat{U}\ket{\psi}\, ,\nonumber
\eea
with $\hat{U}=\exp\lr{-i\o_c t \ahd\ah -i\o_c t \sum\limits_j\s^z_j/2}$. The resulting Hamiltonian now takes the form
\bea
\hat{H}&=&\hat{H}_o+\hat{H}_I\, ,\nonumber\\
\hat{H}_o&=&\frac{1}{2} \Delta \sum\limits_{j=1}^N \s^z_j + \delta \ahd\ah+ A\sum \limits _{j=1}^N \lr{ \s^+_j+ \s^-_j}\, ,\nonumber\\
\hat{H}_I&=&  \sum\limits_{j=1}^N g_j\lr{\s^+_j \ah+\s^-_j \ahd}\, ,\nonumber
\eea
with  $\Delta=\O-\o_c$ and  $\delta=\o-\o_c$. The interaction picture Hamiltonian $\mathcal{V}=e^{-i \hat{H}_o t}\hat{H}_I e^{i \hat{H}_o t}$ upon setting the qubits in resonance with the classical field $\Delta=0$  yields
{\small
\bea
\mathcal{V}= \sum\limits_{j=1}^N
 \frac{1}{2}g_j (\out{+_j}{+_j}-\out{-_j}{-_j}+e^{2 i A t} \out{+_j}{-_j}\nonumber\\
~~~~~~-e^{-2 i A t}\out{-_j}{+_j} )\ah e^{-i \delta t} + h.c. \, ,\nonumber
\eea}
where $\ket{\pm_j}=\frac{1}{\sqrt{2}}\lr{\ket{e_j}\pm \ket{g_j}}$ are the eigenstates of the Pauli $\s^x$ matrix in the $j^{\rm th}$ qubit space.
Disregarding the
terms and redefining $\frac{1}{2}g_j\to g_j$, we obtain
\bea\label{multipleHs}
\mathcal{V}= \sum\limits_{j=i} g_j \s_j^x \lr{\ah e^{-i \delta t}+\ahd e^{i \delta t}}=\sum\limits_{j=i}\mathcal{V}_j \, ,
\eea

{\em Master equation, one qubit}.
Let us first focus on an interaction between a single qubit $N=1$ and a coherent state of the cavity $\a$.
The evolution of this system in a dissipative cavity is driven by the Lindblad-type master equation, 
\begin{equation}\label{lindblad}
\der{\r}{t}=\frac{1}{i\hbar}\com{\hat{\mathcal{V}}}{\r}+\k \mathcal{D}\lr{\r}\, ,
\end{equation}
where $\mathcal{D}\lr{\r}=2\ah\r\ahd- \ahd\ah \r-\r \ahd\ah  \equiv 2 \mathcal{M}\lr{\r}-\mathcal{R}\lr{\r}-\mathcal{L}\lr{\r}$ is the so-called dissipation operator and $\k$ represents the cavity decay rate. 

Using the interaction picture  Hamiltonian  and expressing the qubit density matrix in the $\ket{\pm}$ basis as $\r\lr{t}=p_{kl}\lr{t}\out{k}{l}$, $k,l = \pm$,
one can write the equations of motion (\ref{lindblad}) for individual density matrix entries,
\bea
\dot{p}_{++}\lr{t} \out{\a}{\a}&=& -ig\com{ \ah e^{-i \delta t}+\ahd e^{i \delta t} }{\out{\a}{\a}}p_{++}\lr{t}\nonumber\\ &~&~~~~~~~~+p_{++}\lr{t}\k\mathcal{D}\lr{\out{\a}{\a}}\label{ppp}\, ,\\
\dot{p}_{+-}\lr{t} \out{\a}{\a}&=& -ig\acom{ \ah e^{-i \delta t}+\ahd e^{i \delta t} }{\out{\a}{\a}}p_{+-}\lr{t}\nonumber\\&~&~~~~~~~~+p_{+-}\lr{t}\k\mathcal{D}\lr{\out{\a}{\a}}\label{ppm}\, .
\eea
Here $\com{\cdot}{\cdot}$ $\lr{\acom{\cdot}{\cdot}}$   denote (anti-)commutator brackets. Additionally equations for $p_{--}\lr{t}$ and $p_{-+}\lr{t}$  are obtained by substituting $g\to-g$ in Eqs. (\ref{ppp}) and (\ref{ppm}) respectively.

These decoupled equations can be solved using the superoperator method  \cite{ZhangXu, ZhangXu2} assuming that the cavity is initiated in the coherent state $\ket{\a}$
\bea\label{sol1qubit}
p_{++}\lr{t} &=& e^{g\mathcal{A}/\delta+\k\mathcal{D} t}\out{\a}{\a} p_{++}\lr{0} \, , \\
p_{+-}\lr{t} &=& e^{\left(g\mathcal{B}/ \delta+\k\mathcal{D} t\right)}\out{\a}{\a} p_{+-}\lr{0} \, . \nonumber
\eea
where we define
\bea
\mathcal{A}&=&c_{-}\lr{t} \lr{\cdot}\ah  -c_{-}\lr{t}\ah\lr{\cdot} +c_{+}\lr{t}\ahd\lr{\cdot}-c_{+}\lr{t}\lr{\cdot}\ahd\, , \nonumber\\
\mathcal{B}&=&\mathcal{X}+\mathcal{Y}\, ,\nonumber\\
\mathcal{X}&=&2\lr{c_{-}\lr{t}\ah\lr{\cdot}-c_{+}\lr{t}\lr{\cdot}\ahd}\, ,\nonumber\\
\mathcal{Y}&=&c_{-}\lr{t}\lr{\cdot}\ah -c_{-}\lr{t}\ah\lr{\cdot} -c_{+}\lr{t}\ahd\lr{\cdot}+c_{+}\lr{t}\lr{\cdot}\ahd\, ,\nonumber\\
c_{\pm}\lr{t}&=&\pm i \delta \int_0^t e^{\pm i\delta t'}{\rm d} t' =e^{\pm i\delta t}-1\, .\nonumber
\eea
Next, using the commutation relations
\bea
\com{\ah{\cdot}}{\ahd{\cdot}}=1 \, , ~~~~\com{{\cdot}\ah }{{\cdot}\ahd }=-1 \, ,\nonumber
\eea
we obtain
\bea
\com{\mathcal{D}}{\mathcal{A}}&=&-\mathcal{A}\, ,~~~~\com{\mathcal{D}}{\mathcal{X}}=\mathcal{X}\, ,~~~~\com{\mathcal{D}}{\mathcal{Y}}=-\mathcal{Y}\, ,\nonumber\\
\com{\mathcal{A}}{\mathcal{B}}&=&0\, ,~~~~~~~ \com{\mathcal{X}}{\mathcal{Y}}=-8\lr{1-\cos\delta t}\, ,\nonumber
\eea
and  later use the Baker–-Campbell–-Hausdorff formula to decompose (\ref{sol1qubit})  into
{\small
\bea
p_{++}\lr{t} &=& e^{\frac{g}{ \delta \k t}\lr{1-e^{-\k t}}\mathcal{A}}e^{\k\mathcal{D} t}p_{++}\lr{0}\out{\a}{\a}  \nonumber\\
&=& p_{++}\lr{0}\out{\a e^{-\k t}-f\lr{t}c_{+}\lr{t} }{\a e^{-\k t}-f\lr{t}c_{+}\lr{t}} \, , \nonumber\\
p_{+-}\lr{t}  &=&e^{h_1\lr{t}}e^{\frac{g}{ \delta t \k}\lr{1-e^{-\k t}}\mathcal{Y}} e^{\k\mathcal{D} t}e^{-\frac{g}{ \delta t \k}\lr{1-e^{-\k t}}\mathcal{X}}\out{\a}{\a} p_{+-}\lr{0}  \nonumber\\
&=&e^{h_1\lr{t}+h_2\lr{t}}   p_{+-}\lr{0}\nonumber\\
&\times&  \out{\a e^{-\k t}+ f\lr{t}c_{+}\lr{t}}{\a e^{-\k t}-f\lr{t}c_{+}\lr{t}} \, . \nonumber
\eea}
In the above we defined
{\small
\bea
f\lr{t}&=& \frac{g}{\delta t \k }\lr{1-e^{-\k t}}\, ,\nonumber\\
h_1\lr{t}&=&-\lr{1- \cos \delta t}\lr{\frac{8g^2 }{\delta^2 t^2 \k ^2}\lr{e^{-\k t}-1+\k t}+4f^2}\, , \nonumber\\
h_2\lr{t}&=&-2if\lr{2-e^{-\k t}} \lr{{\rm Im}\lr{\a}\lr{\cos\delta t -1}-{\rm Re}\lr{\a}\sin\delta t} \, . \nonumber
\eea}

Extending this treatment to two qubits in a single cavity requires taking another copy of the interaction Hamiltonian (\ref{multipleHs}).
The only difference  is that now there will be more Hamiltonians $\hat{\mathcal{V}}_i$, acting separately on  different qubit states  and  jointly on the same cavity state.
The solutions are obtained analogously to a single qubit case (see Appendix). 

\begin{figure}
	\centering
		\includegraphics[width=8cm]{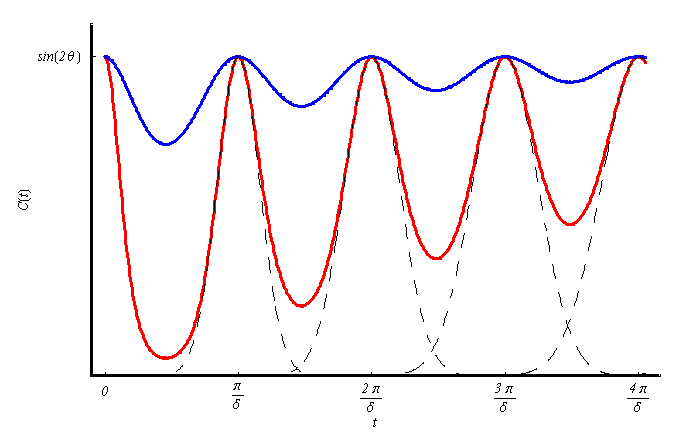}
		\caption{Concurrence for the Bell $\Psi$ (red) and Bell $\Phi$ (blue) states. We see that the amplitude of variation  is significantly smaller and the entanglement recovery speed are greater in case of the later ones. Dashed lines are Gaussians with standard deviations given by equation (\ref{stddev}). Plots made for $g_1=1$, $g_2=0.5$, and $\k=1$. }
	\label{fig:gr}
\end{figure}

\emph{Entanglement Evolution}. Using the approach first proposed in \cite{PhysRevLett.80.2245}, we can now quantify the degree of entanglement of a $2\times 2$ system by means of concurrence defined as
$$
C={\rm max} \lr{0,\sqrt{\lambda_1}-\sqrt{\lambda_2}-\sqrt{\lambda_3}-\sqrt{\lambda_4}}\, ,
$$
where $\lambda_i$ are the descending eigenvalues 
of the real matrix $R=\lr{\s_y\otimes\s_y}\r^*\lr{\s_y\otimes\s_y}\r$. We will assume that the qubit pair is initialised in either of the  generalised set of  Bell states
\bea
\Psi &=&\cos\theta \ket{++}+e^{i\phi}\sin\theta\ket{--}\, ,\nonumber\\
\Phi &=&\cos\theta\ket{+-}+e^{i\phi}\sin\theta\ket{-+}\, ,\nonumber
\eea
and afterward it is evolving according to the dynamics given by (\ref{lindblad}). As a result, the concurrence is a non-trivial function of time
\bea\label{concurrence2}
C=\sin2\theta e^{-8\lr{1-\cos\delta t} (g_1\pm g_2)^2 \left(\k  t+e^{-\k  t}-1\right)/\lr{\k ^2 \delta ^2 t^2}} \,,
\eea
where   the upper (lower) sign is used to denote the entanglement evolution of the $\Psi$ $\lr{\Phi}$ states. The graph for the evolution of both of these is plotted in Fig. \ref{fig:gr}.
By approximating each of the peaks with a Gaussian we find that every consecutive maximum will have the form
{\small
$$
C_n\lr{t}= \sin 2\theta \exp\lr{-\frac{\lr{t-2\pi n/\delta}^2}{\tau_n^2}}\, ,
$$}
where we used the standard deviation $\tau_n$ to be a measure of every consecutive revival time given  by
\begin{equation}\label{stddev}
\tau_n=\frac{2 \sqrt{2} \k  n \pi }{g_1\pm g_2}  \lr{-2 \k  \pi \delta n +\delta^2\lr{1-e^{-\frac{2 \k  n \pi }{\delta }}} }^{-\frac{1}{2}}\, .
\end{equation}

\begin{figure}
	\centering
		\includegraphics[width=8cm]{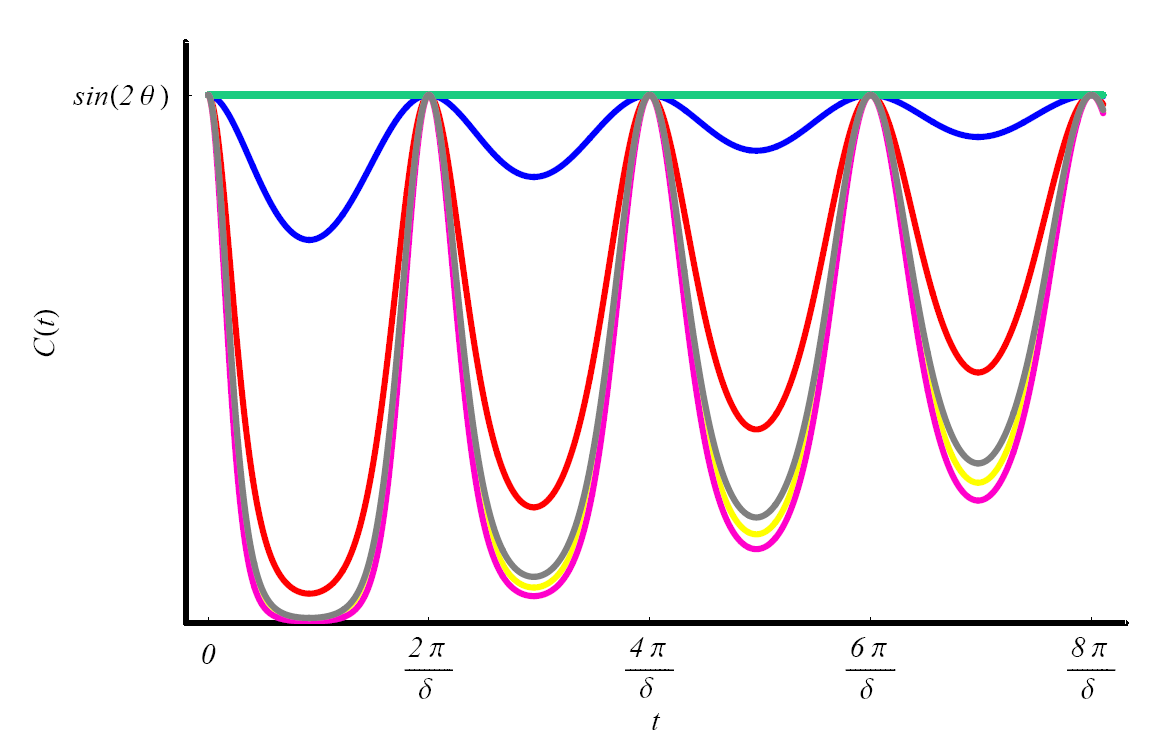}
		\caption{Concurrence for the Bell $\Psi$ (red$^{(i)}$, yellow$^{(ii)}$ and pink$^{(iii)}$) and Bell $\Phi$ ( blue$^{(i)}$,   green$^{(ii)}$ and  gray$^{(iii)}$) states. Plots made for $\k=1$ (i) $g_1=1$, $g_2=0.5$, (ii) $g_1=1$, $g_2=1$ and (iii) $g_1=2$, $g_2=0.1$. If the coupling strengths are the same (ii) $\Phi$ will experience on changes. If the relative coupling strength is large (iii) the concurrences for $\Phi$ and $\Psi$ are very similar. }
	\label{fig:gr2}
\end{figure}

\begin{figure}
	\centering
		\includegraphics[width=8cm]{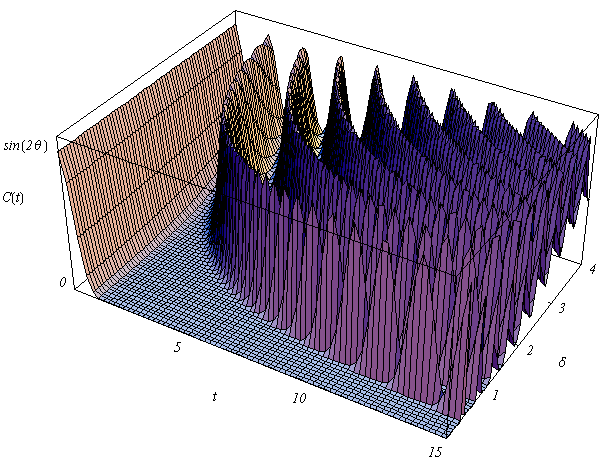} 
		\caption{A 3D plot of concurrence of $\Psi$ as a function of time and detuning. For detunings $\delta\ll 1$ we observe the entanglement sudden death behaviour already noted by \cite{Bina2008}. For increasing values of detuning, the concurrence function reaches the steady state maximum value quicker. Plots made for $g_1=1$, $g_2=0.5$, and $\k=0.1$.}
	\label{fig:gr5}
\end{figure}

\begin{figure}
	\centering 
		\includegraphics[width=8cm]{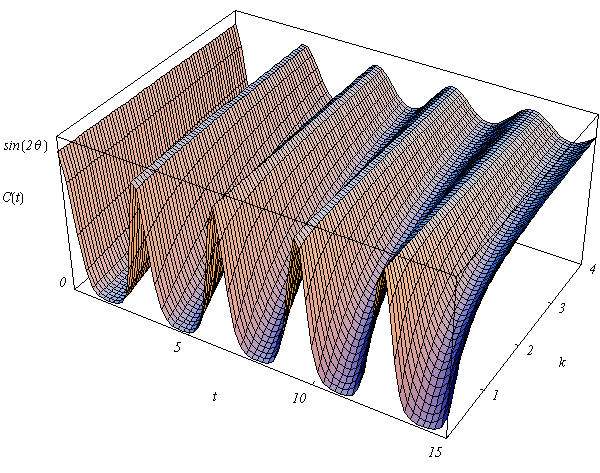}
		\caption{A 3D plot of concurrence of $\Psi$ as a function of time and cavity decay rate. For decay rates $\k\ll 1$ we observe that less entanglement is lost and its asymptotic recovery is quicker.  Plots made for $g_1=1$, $g_2=0.5$, and $\delta=2$.}
	\label{fig:gr6}
\end{figure}

Eq. (\ref{concurrence2}) displays a number of striking properties.  Firstly, after an initial sharp decrease the concurrence periodically  recovers its initial value $\sin 2\theta$  never exceeding it throughout. This confirms the previous result that qubit-qubit  entanglement enhancement is not possible in this system \cite{Bina2008, Bina2009}. Secondly,   the entanglement exhibits oscillatory behaviour showing  periodic  revivals at $\delta t=2 n \pi$, with the revival time intervals $\tau_n\to \infty$ as $n,t\to \infty$. Thirdly, the greater the rate of cavity decay $\k$ (Figure \ref{fig:gr5}) or the degree of detuning $\delta$ (Figure \ref{fig:gr6}), the quicker is the recovery of the initially entangled state. The reason for that is that with greater $\k$  the cavity eigenmode field deplets quicker and so  the chance for qubits to interact with the quantum field decreases. This effect is enhanced if the qubits are detuned from the quantum eigenmode inhibiting interaction. As a result both of these effects lead to a decreased opportunity of disentanglement. Moreover, unlike Refs. \cite{Bina2008, Bina2009, Bina2010}, we have chosen to work with  an arbitrary initial coherent state amplitude $\a\lr{0}\neq 0$ to  observe that its value plays no role in the qubit-qubit entanglement evolution, thus making this result universal for all cavities.

Finally we also find, in line with  Refs. \cite{Bina2009,Bina2010}, that qubits initialised to different  Bell states respond differently in this system. In Refs. \cite{Bina2009, Bina2010} the authors claimed that the concurrence of  the $\Phi$ type states is unaffected by cavity dissipation.
We find that this is only true if the qubits are equally coupled to the cavity vacuum field. As a result all Bell states formed with unequally coupled qubits will decay and be revived depending on the values of $\delta$ and $\k$, however the $\Phi$ states will do it at a slower rate and the value of concurrence will drop to a lesser extent (see Fig. \ref{fig:gr}).

To simplify calculations, we have chosen the regime $A\gg \o,\delta \gg g \gg \gamma$. These conditions can be realized in two types of solid-state qubits. In
a superconducting qubit coupled to a microwave cavity \cite{PhysRevA.80.043840}, one can achieve $\o \approx 5$ GHz, $g \approx 100$ MHz, and $\gamma \approx 1$ MHz. The critical parameter here is the qubit dissipation $\gamma$. We do expect however that the condition $g \gg \gamma$ can be relaxed without qualitatively affecting the results. Another system is an NV center in diamond strongly coupled to an optical cavity \cite{2010arXiv1005.4428R}, used as a spin qubit. For this realization, the coupling strength is the crucial parameter to observe the entanglement revival.


\emph{Conclusions}.
We have shown that a pair of qubits detuned from the quantum field, initially show a sudden decrease of entanglement to later gradually recover its initially entangled state. Additionally we show that no Bell states are strictly protected, however they can be better secured by skillful adjustment of the detuning parameter.

This study can be extended by considering the dynamics of a tripartite system composed of any Bell state and the coherent state of the cavity. Here one can consider the evolution of entanglement between any selected pair of subsystems and study entanglement creation between the cavity and the qubits during the qubit-qubit disentanglement phase. These results will be published elsewhere \cite{me}.

Moreover, by virtue of extendibility of this model to an arbitrary number of qubits as well as cavities, using this framework one could study multipartite entanglement and how, depending on the conditions and parameters choice, entanglement could be exchanged or transferred between  different subsystems. Additionally, already with three qubits in the cavity one could try to find more differences in evolution between two maximally entangled classes:  the GHZ and the W states.
\vspace{0.3cm}

The authors wish to thank Antonio Borras Lopez, Pol Forn-Diaz and Toeno van der Sar for useful discussions. This work was supported by the Netherlands Foundation for Fundamental Research
on Matter (FOM).

\bibliography{entangelment_periodic_revival3}

\appendix
\section{Appendix}
Here we present the equation and solutions to Eq.(\ref{lindblad}) for a two qubit case.

The density matrix of the two qubits states is labeled by
 $p_{ij;kl, \alpha} (t)$, where $i,j=\pm$ and $k,l=\pm$ refer to the first and second qubits, respectively. 
Additionally in every entry there is a distinct coherent state density operator $\out{\a}{\a}_{ij;kl}$ present. Equations for individual matrix components read 
\begin{widetext}
\bea
\dot{p}_{++;++}  \out{\a}{\a}&=& p_{++;++} \lr{\frac{g_1+g_2}{i\hbar}\com{ \ah e^{-i \delta t}+\ahd e^{i \delta t}}{\out{\a}{\a}}+  \k\mathcal{D}\lr{\out{\a}{\a}}}\, ,\nonumber\\
\dot{p}_{++;+-}  \out{\a}{\a}&=& p_{++;+-}\lr{ \frac{g_1 }{i\hbar}\com{ \ah e^{-i \delta t}+\ahd e^{i \delta t}}{\out{\a}{\a}}+  \frac{g_2 }{i\hbar}\acom{ \ah e^{-i \delta t}+\ahd e^{i \delta t}}{\out{\a}{\a}}+  \k\mathcal{D}\lr{\out{\a}{\a}}}\, ,\nonumber\\
\dot{p}_{++;-+}  \out{\a}{\a}&=& p_{++;-+}\lr{ \frac{g_1 }{i\hbar}\com{ \ah e^{-i \delta t}+\ahd e^{i \delta t}}{\out{\a}{\a}}-  \frac{g_2 }{i\hbar}\acom{ \ah e^{-i \delta t}+\ahd e^{i \delta t}}{\out{\a}{\a}}+  \k\mathcal{D}\lr{\out{\a}{\a}}}\, ,\nonumber\\
\dot{p}_{++;--}  \out{\a}{\a}&=& p_{++;--}\lr{ \frac{g_1+g_2}{i\hbar}\acom{ \ah e^{-i \delta t}+\ahd e^{i \delta t}}{\out{\a}{\a}}+ \k\mathcal{D}\lr{\out{\a}{\a}}}\, ,\nonumber\\
\dot{p}_{+-;+-}  \out{\a}{\a}&=& p_{+-;+-}\lr{ \frac{g_1-g_2}{i\hbar}\com{ \ah e^{-i \delta t}+\ahd e^{i \delta t}}{\out{\a}{\a}}+  \k\mathcal{D}\lr{\out{\a}{\a}}}\, ,\nonumber\\
\dot{p}_{+-;-+}  \out{\a}{\a}&=& p_{+-;-+}\lr{ \frac{g_1-g_2}{i\hbar}\acom{ \ah e^{-i \delta t}+\ahd e^{i \delta t}}{\out{\a}{\a}}+ \k\mathcal{D}\lr{\out{\a}{\a}}}\, ,\nonumber\\
\dot{p}_{+-;--}  \out{\a}{\a}&=& p_{+-;--}\lr{ \frac{g_1}{i\hbar}\acom{ \ah e^{-i \delta t}+\ahd e^{i \delta t}}{\out{\a}{\a}}- \frac{g_2 }{i\hbar}\com{ \ah e^{-i \delta t}+\ahd e^{i \delta t}}{\out{\a}{\a}}+  \k\mathcal{D}\lr{\out{\a}{\a}}}\, ,\nonumber\\
\dot{p}_{-+;-+}  \out{\a}{\a}&=& p_{-+;-+}\lr{ \frac{-g_1+g_2}{i\hbar}\com{ \ah e^{-i \delta t}+\ahd e^{i \delta t}}{\out{\a}{\a}}+  \k\mathcal{D}\lr{\out{\a}{\a}}}\, ,\nonumber\\
\dot{p}_{-+;--}  \out{\a}{\a}&=& p_{-+;--}\lr{ \frac{-g_1}{i\hbar}\com{ \ah e^{-i \delta t}+\ahd e^{i \delta t}}{\out{\a}{\a}}+  \frac{g_2 }{i\hbar}\acom{ \ah e^{-i \delta t}+\ahd e^{i \delta t}}{\out{\a}{\a}}+  \k\mathcal{D}\lr{\out{\a}{\a}}}\, ,\nonumber\\
\dot{p}_{--;--}  \out{\a}{\a}&=& p_{--;--}\lr{- \frac{g_1+g_2}{i\hbar}\com{ \ah e^{-i \delta t}+\ahd e^{i \delta t}}{\out{\a}{\a}}+ \k\mathcal{D}\lr{\out{\a}{\a}}}\nonumber\, .
\eea 
\end{widetext}
The solutions to these equations 
can be easily obtained by identifying every commutator with an  $\mathcal{A}$ and every anti-commutator with a $\mathcal{B}$ operator. The solutions are again exponents of operators acting on $\out{\a}{\a}$ which can be decomposed using the Baker--Campbell-Hausdorff formula. By first using the fact that operators $\mathcal{A}$ and $\mathcal{B}$ commute to decompose $\mathcal{A}$ from $\frac{g}{\delta}\mathcal{B}+\k\mathcal{D} t$ we later follow the single qubit case steps to obtain
\begin{widetext}
\bea
p_{++;++}\lr{t}    &=&  p_{++;++}\lr{0}\out{\a e^{-kt}-  g_1 f c_{+}-  g_2 f c_{+}}{\a e^{-kt}-  g_1 f c_{+}-  g_2 f c_{+}} \, ,\nonumber\\
p_{++;+-}\lr{t}    &=&  p_{++;+-}\lr{0} e^{x\lr{g_2,t} } \out{\a e^{-kt}-  g_1 f c_{+}+  g_2 f c_{+}}{\a e^{-kt}-  g_1 f c_{+}-  g_2 f c_{+}}\, ,\nonumber\\
p_{++;-+}\lr{t}    &=&  p_{++;-+}\lr{0} e^{x\lr{g_1,t} } \out{\a e^{-kt}+  g_1 f c_{+}-  g_2 f c_{+}}{\a e^{-kt}-  g_1 f c_{+}-  g_2 f c_{+}}\, ,\nonumber\\
p_{++;--}\lr{t}    &=&  p_{++;--}\lr{0} e^{x\lr{g_1+g_2,t} }\out{\a e^{-kt}+  g_1 f c_{+}+  g_2 f c_{+}}{\a e^{-kt}-  g_1 f c_{+}-  g_2 f c_{+}}\, ,\nonumber\\
p_{+-;+-}\lr{t}    &=&  p_{+-;+-}\lr{0} \out{\a e^{-kt}-  g_1 f c_{+}+  g_2 f c_{+}}{\a e^{-kt}-  g_1 f c_{+}+  g_2 f c_{+}}\, ,\nonumber\\
p_{+-;-+}\lr{t}    &=&  p_{+-;-+}\lr{0} e^{x\lr{g_1-g_2,t}}  \out{\a e^{-kt}+  g_1 f c_{+}-  g_2 f c_{+}}{\a e^{-kt}-  g_1 f c_{+}+  g_2 f c_{+}}\, ,\nonumber\\
p_{+-;--}\lr{t}    &=&  p_{+-;--}\lr{0} e^{x\lr{g_1,t}} \out{\a e^{-kt}+  g_1 f c_{+}+  g_2 f c_{+}}{\a e^{-kt}-  g_1 f c_{+}+  g_2 f c_{+}}\, ,\nonumber\\
p_{-+;-+}\lr{t}    &=&  p_{-+;-+}\lr{0} \out{\a e^{-kt}+  g_1 f c_{+}-  g_2 f c_{+}}{\a e^{-kt}+  g_1 f c_{+}-  g_2 f c_{+}}\, ,\nonumber\\
p_{-+;--}\lr{t}    &=&  p_{-+;--}\lr{0} e^{x\lr{g_2,t} } \out{\a e^{-kt}+  g_1 f c_{+}+  g_2 f c_{+}}{\a e^{-kt}+  g_1 f c_{+}-  g_2 f c_{+}}\, ,\nonumber\\
p_{--;--}\lr{t}    &=&  p_{--;--}\lr{0} \out{\a e^{-kt}+  g_1 f c_{+}+  g_2 f c_{+}}{\a e^{-kt}+  g_1 f c_{+}+  g_2 f c_{+}}\, .\nonumber
\eea
\end{widetext}
where the solutions to the remaining six entries of the density operator are found by Hermitian conjugation and where $f$ and $c_+$ were defined before. Additionally, in the above we defined  
 \begin{widetext}
\bea
x\lr{\xi,t}&=&-\frac{8\xi^2 \lr{1- \cos \delta t}}{\delta^2\k^2 t^2}\lr{e^{-\k t}-1+\k t}+4\xi^2 f^2\lr{\cos \delta t-1}-2i\xi f\lr{2-e^{-\k t}} \lr{{\rm Im}\lr{\a}\lr{\cos\delta t -1}-{\rm Re}\lr{\a}\sin\delta t}   \nonumber\, .
\eea 
\end{widetext}
These solutions can be used to extend the treatment to more then one unequally coupled qubits and to study multipartite entanglement in this system. Tracing out the cavity  leads to a two-qubits  reduced density matrix
\begin{widetext}
{ $$
\r_{q_1,q_2 }=\lr{\begin{array}{cccc}
p_{++;++}\lr{0}   & p_{++;+-}\lr{0}\varrho ^{-}\lr{g_2}    &  p_{++;-+}\lr{0}\varrho ^{-} \lr{g_1}  &p_{++;--}\lr{0} \varrho ^{-} \lr{g_1+g_2}     \\
p_{+-;++}\lr{0}\varrho ^{+}\lr{g_2}         & p_{+-;+-}\lr{0}     &  p_{+-;-+}\lr{0}\varrho ^{-}\lr{g_1-g_2}   & p_{+-;--}\lr{0}\varrho ^{-}\lr{g_1}   \\
p_{-+;++}\lr{0}\varrho ^{+}\lr{g_1}        &p_{-+;+-}\lr{0} \varrho ^{+}\lr{g_1-g_2 }   &  p_{-+;-+}\lr{0} &p_{-+;--}\lr{0} \varrho ^{-}\lr{g_2}     \\
p_{--;++}\lr{0}\varrho ^{+} \lr{g_1+g_2}    & p_{--;+-}\lr{0}\varrho ^{+}\lr{g_1}           &  p_{--;-+}\lr{0}\varrho ^{+}\lr{g_2}                & p_{--;--}\lr{0}
\end{array}}\, ,
$$}
where we  define
\bea
\varrho ^{\pm}\lr{\xi}&=&\exp\lr{-\frac{8\xi^2 \lr{1- \cos \delta t}}{\delta^2 t^2 \k ^2}\lr{e^{-\k t}-1+\k t} }  \exp\lr{-\frac{4i\xi}{\delta t\k } \lr{1-e^{-\k t}}^2 \lr{{\rm Im}\lr{\a}\lr{\cos\delta t -1}-{\rm Re}\lr{\a}\sin\delta t}}\, .   \nonumber
\eea
\end{widetext}
\begin{minipage}{1\linewidth}
Upon imposition of initial conditions this result can be used to study entanglement dynamics by means of Wootters' concurrence.
\end{minipage}
\end{document}